\documentclass[a4paper,preprint,showpacs,amssymb,pre,superscriptaddress]{revtex4}

\usepackage{graphicx}
\usepackage{graphicx}
\usepackage{amsfonts}
\usepackage{amssymb}
\begin{document}

\title{Noise effects in extended chaotic system: study on the
Lorenz'96 model}

\author{J. A. Revelli, M. A. Rodriguez and H. S. Wio}

\affiliation{Instituto de F\'{\i}sica de Cantabria, Universidad de
Cantabria and CSIC \\ E - 39005 Santander, Spain}

\begin{abstract}
We investigate the effects of a time-correlated noise on an extended
chaotic system. The chosen model is the Lorenz'96, a kind of toy
model used for climate studies. The system is subjected to both
temporal and spatiotemporal perturbations. Through the analysis of
the system $'s$ time evolution and its time correlations, we have
obtained numerical evidence for two stochastic resonance-like
behaviors. Such behavior is seen when a generalized signal-to-noise
ratio function are depicted as a function of the external noise
intensity or as function of the system size. The underlying
mechanism seems to be associated to a noise-induced chaos reduction.
The possible relevance of those findings for an optimal climate
prediction are discussed, using an analysis of the noise effects on
the evolution of finite perturbations and errors.
\end{abstract}

\maketitle

\section{Introduction}

It is well known that noise and chaos represent, respectively, two
kinds of essentially different phenomena. The former is induced by
genuine stochastic sources, while the randomness of the later is
pseudo and is deterministic in its origin. The spatiotemporal chaos
is intrinsically irregular in both space and time and represents a
prototype of deterministic randomness. It is interesting to see what
would come about as a result of the interaction between these two
irregularities that are essentially distinct.

As the influence of noises on low-dimensional dynamics systems has
been studied extensively \cite{RMP,mangwio}, much research interest
has nowadays shifted to spatially extended system, a situation that
is apparently much more complicated \cite{sancho}.

In the spatially extended situations, the way in which the noise
takes effect is not obvious and the deterministic description
usually cannot give the right results. It is known that
noise-induced phenomena have come about as a consequence of
nonlinear interaction between the noise and the deterministic
dynamics. The spatiotemporal stochastic resonance are believed to
have potential importance, for instance, in the area of signal and
image processing, pattern formation, social and economical as well
as climate dynamics \cite{RMP,extend2}.

Here we  consider a fully study on the Lorenz'96 model, driven by
two kinds of perturbations, a deterministic perturbation given by
the own chaotic behavior of the model and a stochastic one which we
have assumed as an effective way of including a more realistic
evolution. The relevance of this model rest on the fact that it
represents a simple but still realistic description of some physical
properties of global atmospheric models.

Manifestations of noise on other characteristics of spatiotemporal
chaos such as Lyapunov exponents and dimensions have not been
considered. The results presented here provided a first step in
order to explore the possibilities of complex dynamics coming out
from the interaction between chaos and noise clearly. Further
investigation along this line is desirable.

This work is organized as follows: in section II we describe the
Lorenz'96 model assuming that its evolution is governed by both a
deterministic and a stochastic processes. In section III we present
and discuss numerical simulations of the Lorenz'96 equation,
describing  qualitatively the interaction of the real noise and the
deterministic noise on the time evolution of the system. In section
IV we discuss the important problem of the perturbations and errors
in the Lorenz'96 evolution. Finally in section V we present the
conclusions of our work as well as possible implications and the
relevance of this study on the actual climate evolution.

\section{The Lorenz'96 model}

The equations corresponding to the Lorenz'96 model are
\begin{equation}
\label{Lorenz96_1} \dot{x}_{j}(t) = - x_{j - 1} (x_{j - 2} - x_{j +
1}) - x_{j} + F,
\end{equation}
where $\dot{x}_{j}$ indicates the time derivative of $x_{j}$
\begin{equation}
\label{Lorenz96_2} F_{j}(t) = F_{med} + \Psi_{j}(t),
\end{equation}
with $\Psi_{j}(t)$ a dichotomic process. That is, $\Psi_{j}(t)$
adopts the values $\pm \Delta$ with a transition rate $\gamma$: each
state changes according to the waiting time distribution $\varphi_i
(t) \sim e^{- \gamma t}$. The noise intensity for this process is
defined through $\xi = \frac{\Delta^2}{2 \gamma}$. In this work we
have supposed that  the system is subjected to a spatiotemporal
perturbation as well as a temporal one. The first perturbation is
achieved when $F$ depends on both $j$ and $t$ variables, meanwhile
for temporal perturbation the $F$ function only depends on $t$. In
order to simulate a scalar meteorological quantity extended around a
latitude circle, we consider  periodic boundary conditions $x_0 =
x_N$, $x_{-1} = x_{N-1}$.

As indicated before, the Lorenz'96 model has been heuristically
formulated as the simplest way to take into account certain
properties of global atmospheric models. The terms included in the
equation intend to simulate advection, dissipation, and forcing
respectively. In contrast with other toy models used in the analysis
of extended chaotic systems and based on coupled map lattices, the
Lorenz'96 model exhibits extended chaos when the $F$ parameter
exceeds a determinate threshold value ($F>9/8$) with a spatial
structure in the form of moving waves. The length of these waves is
close to $5$ spatial units. It is worth noting that the system has
scaled variables with unit coefficients, hence the time unit is the
dissipative decay time. In addition we adjust the value of the
parameter $F$ to give a reasonable signal to noise ratio (Lorenz
considered $F=8$), so the model could be most adequate to perform
basic studies of predictability.

\subsection{System Response}

As a measure of the SR system's response we have used the
\textit{signal-to-noise ratio} (SNR) \cite{RMP}. To obtain the SNR
we need to previously evaluate $S(\omega)$, the power spectral
density (psd), defined as the Fourier transform of the correlation
function \cite{vK,Gar}
\begin{equation} \label{lor3}
S (\omega) = \int _{-\infty}^{\infty} e^{i\omega \tau} \langle
x_{j}(0) x_{j}(\tau) \rangle \, d\tau ,
\end{equation}
where $\langle \,\,\, \rangle$ indicates the average over
realizations. As we have periodic boundary conditions simulating a
closed system, $\langle x_{j}(0) x_{j}(\tau) \rangle$ has a
homogeneous spatial behavior. Hence, it is enough to analyze the
response in a single site.

We consider two forms of SNR. In one hand the usual SNR measure at
the resonant frequency $\omega _{o}$ (that is, in fact, at the
frequency associated to the highest peak in $S(\omega)$) is
\begin{equation} \label{lor4}
SNR = \frac{\int ^{\omega _{o} + \sigma}_{\omega _{o} - \sigma} d
\varpi S(\varpi)}{\int ^{\omega _{o} + \sigma}_{\omega _{o} -
\sigma} d  \varpi S_{back}(\varpi)},
\end{equation}
where $2 \sigma$ is a very small range around the resonant frequency
$\omega _{o}$, and $S_{back}(\omega)$ corresponds to the background
psd. On the other hand we consider a global form of the SNR
($SNR_{glob}$) defined through
\begin{equation} \label{lor5}
SNR_{glob} = \frac{\int ^{\omega_{max}}_{\omega_{min}} d \omega
S(\omega)}{\int ^{\omega_{max}}_{\omega_{min}} d \omega
S_{back}(\omega)},
\end{equation}
where $\omega_{min}$ and $\omega_{max}$ define the frequency range
where $S(\omega)$ has a reach peak structure (with several
\textit{resonant} frequencies).

\section{Stochastic resonance-like effects}

\subsection{System Time Evolution}

In this section we present numerical simulations of a Lorenz '96
system subjected to both temporal and spatiotemporal noise
perturbations. The typical numbers we have used in our simulations
are: averages over $10^3$ histories, and $\sim 10^4$ simulation time
steps (within the stationary regime, see later).

\begin{figure}
\centering
\resizebox{.48\columnwidth}{!}{\includegraphics{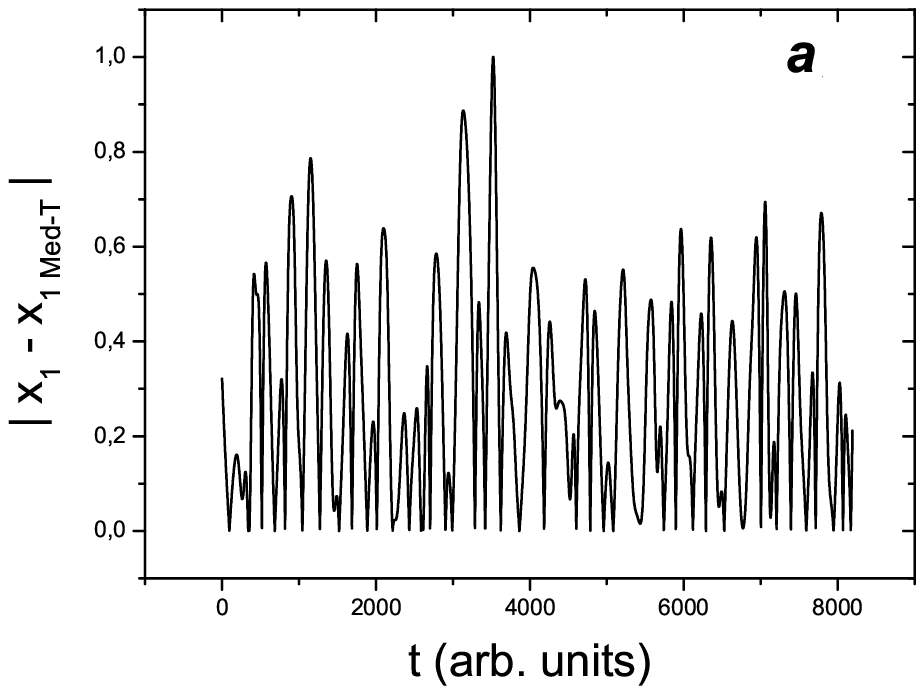}}
\resizebox{.48\columnwidth}{!}{\includegraphics{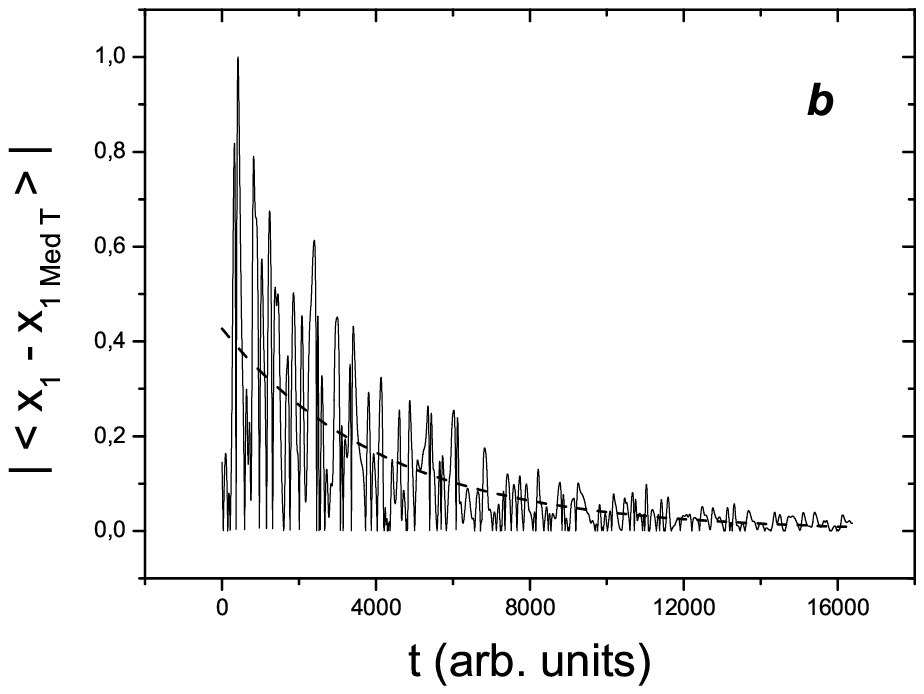}}
\caption{System parameters $N=128$, $F_{Med}=5$. (a) Time evolution
without noise.  (b) Time evolution with noise $\Delta=1$,
$\gamma=0.05$. \label{Fig.1}}
\end{figure}

We have analyzed the typical behavior of trajectories as $x_1(t) -
x_{med-T}$, where $x_{med-T}$ is the time average. When the Lorenz
'96 system evolves without external noise ($F_j(t)$ is constant),
the time evolution shows a random-like behavior. As can be seen in
Fig. \ref{Fig.1}-a  where we describe the $|x_1 - x_{1 Med T}|$, the
main feature is that the amplitude of the oscillator is almost
constant over all the time. If the system is subject to a true
random force, described like in Eq.[\ref{Lorenz96_1}] as shown in
Fig. \ref{Fig.1} - b, then the temporal oscillator response decays,
that is the interaction between the intrinsic evolution and the
external noise produces  dissipation on the system. Hence the time
evolution of the system consists of a transitory regime and a
stationary one.

\begin{figure}
\centering
\resizebox{.48\columnwidth}{!}{\includegraphics{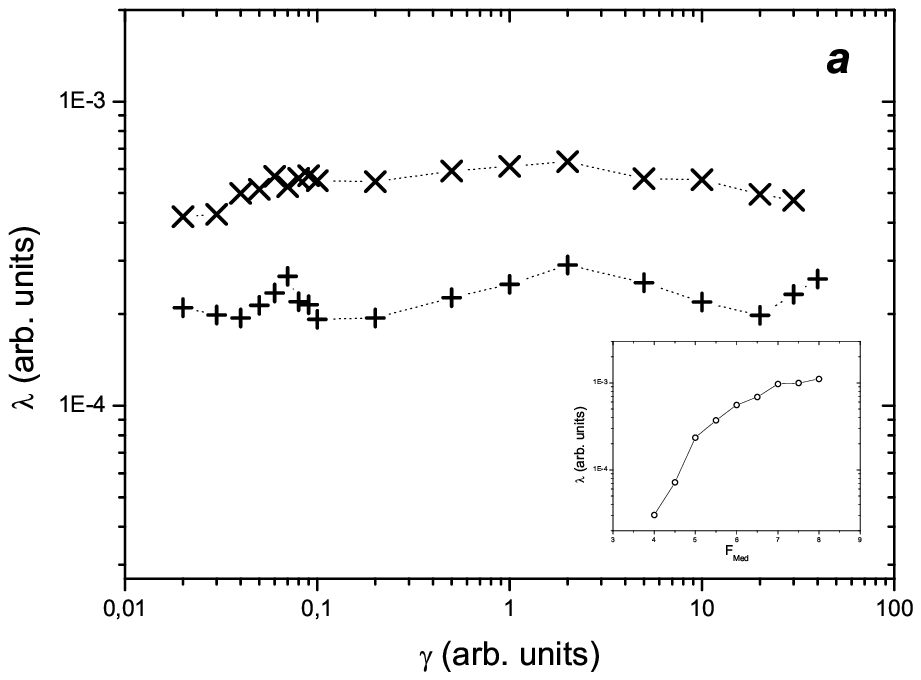}}
\resizebox{.48\columnwidth}{!}{\includegraphics{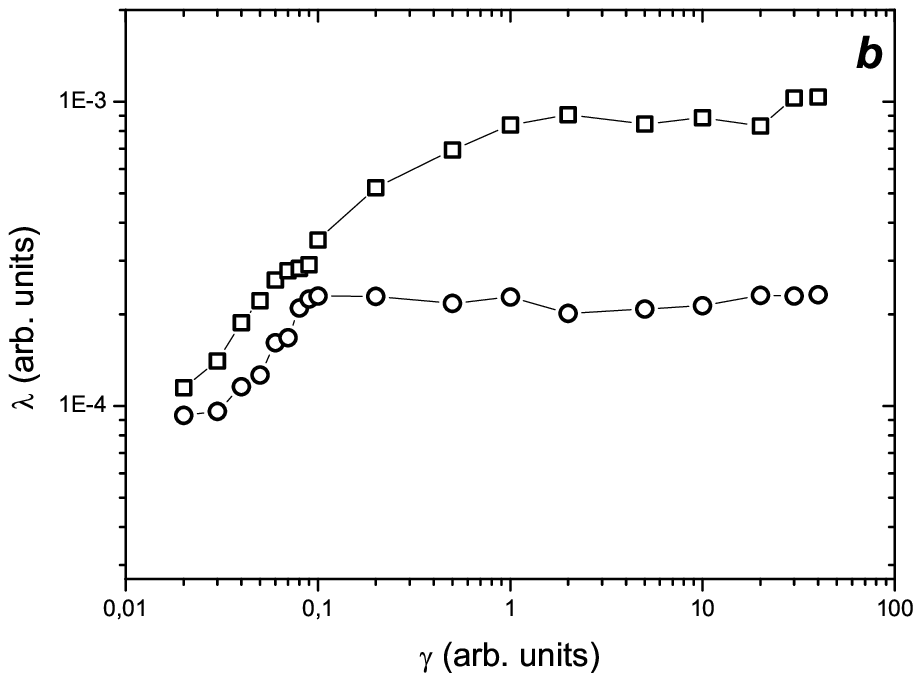}}
\caption{(a) Spatiotemporal evolution. Decay rate vs. transition
rate $\times F_{Med}=6$, $+ F_{Med}=5$, $N=64$ Insert: Decay rate
vs. $F_{Med}$ (b) Temporal evolution. $F_{Med}=6$ circles,
$F_{Med}=5$ squares. \label{Fig.2}}
\end{figure}

We have assumed that this decay can be adjust by an exponential law
($|x_1 - x_{1 \,\ Med \,\ T}| \sim \exp(- \lambda t)$). Figures
\ref{Fig.2}-a depicts  the $\lambda$ dependence on the transition
rate $\gamma$ for a spatiotemporal evolution. The figure shows the
weak dependence of $\lambda$ on $\gamma$ for two $F_{Med}$
parameters. In one hand the temporal decay parameter depends on the
$F_{med}$ parameter as can be seen in the insert of the Fig.
\ref{Fig.2}-a. On the other hand there exists a clear dependence of
$\lambda$ when the system is subject to only temporal perturbation
(see Fig. \ref{Fig.2} - b). Two regions can be observed, firstly a
quasi linear grow for a low rate transitions and a saturation regime
for large $\gamma$. It is worth remarking here that this saturation
regime is not the same than that for a spatiotemporal perturbation.
The indicated time decay is important as we are interested in
studying the noise effects on the stationary regime.

\subsection{\label{resu}Resonant-like behavior}

\begin{figure}
\centering
\resizebox{.48\columnwidth}{!}{\includegraphics{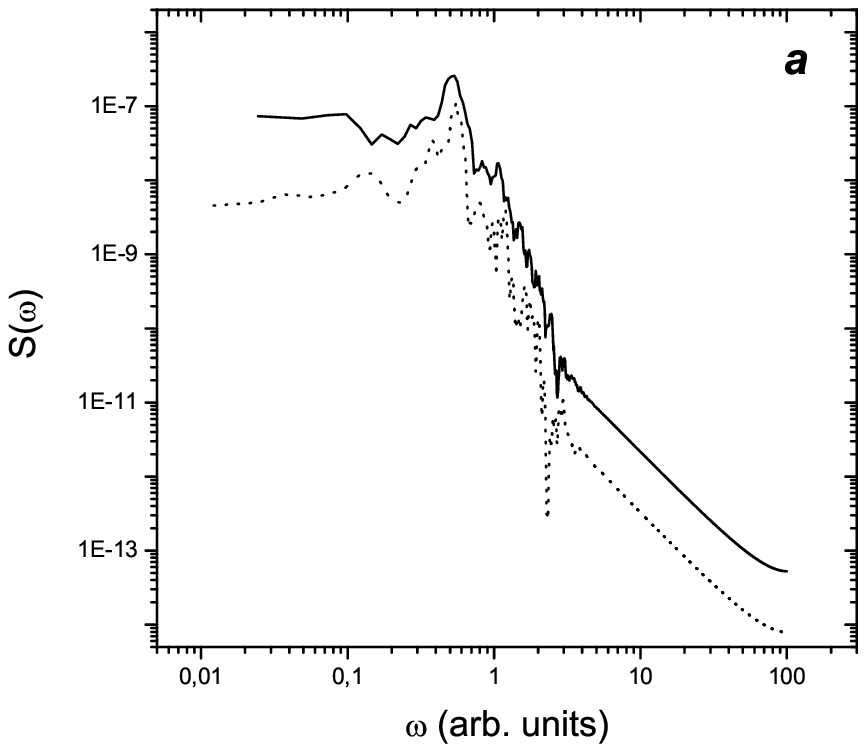}}
\resizebox{.48\columnwidth}{!}{\includegraphics{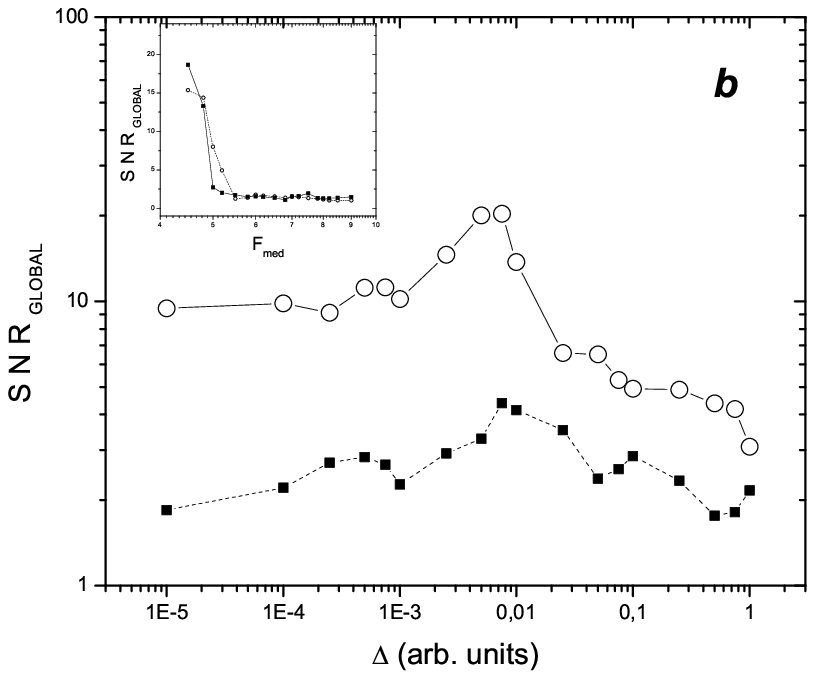}}
\caption{(a) Power spectral density. Solid line: Lorenz $´96$
without noise. Dash line: Lorenz $´96$  evolution with $\Delta=0.1$,
$F_{Med}=5$, $N=128$. (b) Signal to noise ratio for $F_{Med}=5$
circles, $F_{Med}=6$ squares. Insert: signal to noise ratio as a
function of $F_{Med}$. \label{Fig.3}}
\end{figure}

Figure \ref{Fig.3}-a shows the power spectrum density (psd) for two
cases, the Lorenz'96 model without noise (continuous line) and when
subject to a dichotomic noise (dash line). From the figure it is
apparent that the simultaneous action of both deterministic and
stochastic noises induces a background reduction. The consequence of
this effect can be appreciated in the Fig. \ref{Fig.3} - b where it
is possible to anticipate, and identify, the existence of a
resonance-like behavior when the global signal to noise
($SNR_{Global}$) is depicted as a function of the noise intensity
($\Delta$). Indeed the response for $F_{Med}=5$ is better than for
$F_{Med}=6$. The insert of the figure shows the $SNR_{Global}$
response as a function of the $F_{Med}$ parameter for two different
noise intensities. We can see that the response is high for low
values of $F_{Med}$ (low developed chaos) and that is almost
constant for large values of $F_{Med}$ (well developed chaos).

\begin{figure}
\centering
\resizebox{.48\columnwidth}{!}{\includegraphics{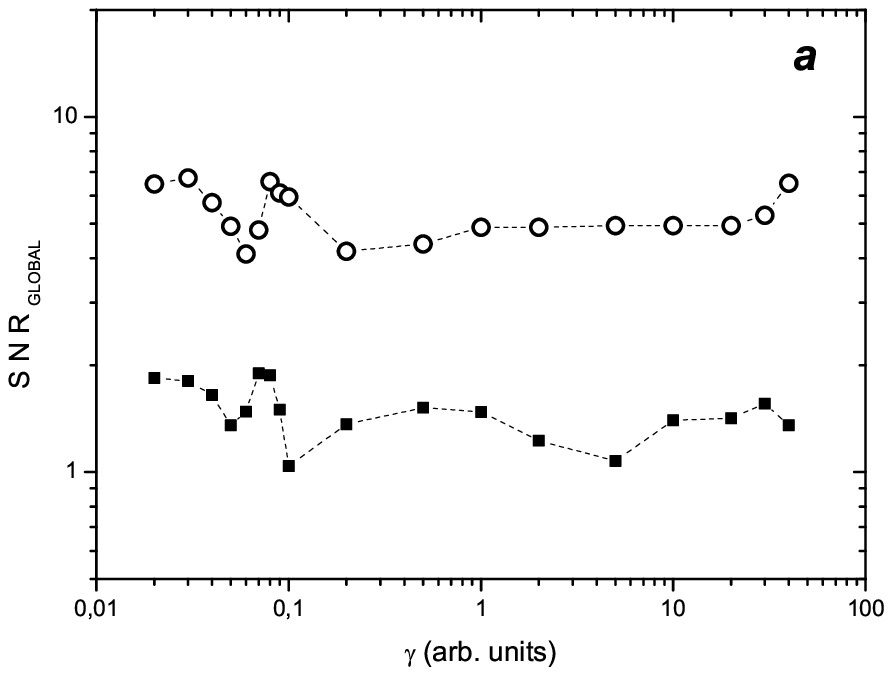}}
\resizebox{.48\columnwidth}{!}{\includegraphics{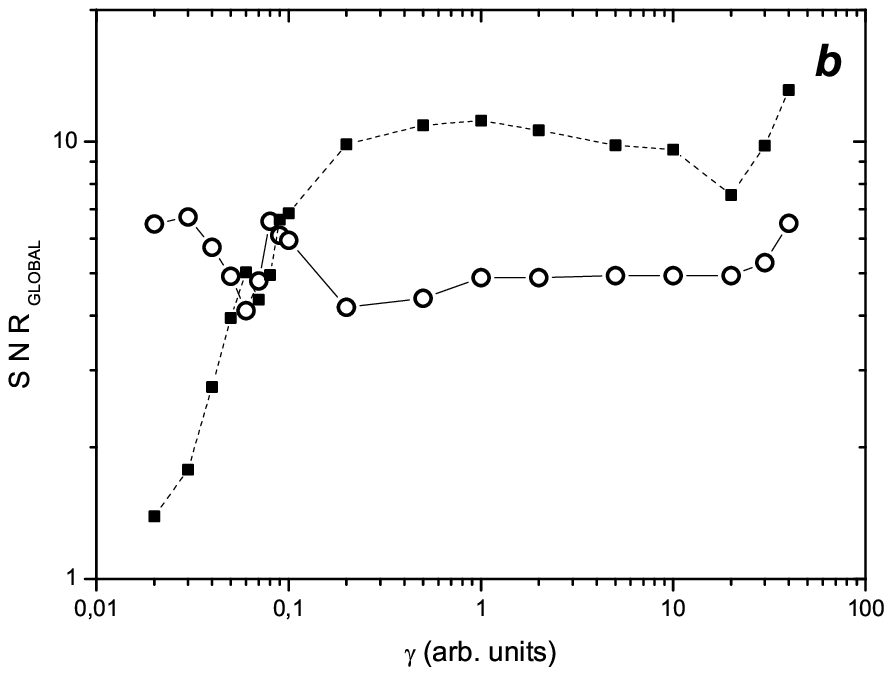}}
\caption{(a) Spatiotemporal evolution. Signal to noise ratio as a
function of the transition rate. $F_{Med}=5$ circles, $F_{Med}=6$
squares. (b)Spatiotemporal (circle) and temporal (square) SNR for a
$N=64$ system, with $\Delta=0.1$. \label{Fig.4}}
\end{figure}

We have also studied the response dependence on the transition rate
$\gamma$. Figure \ref{Fig.4}-a (spatiotemporal noise for two
$F_{Med}$) and Fig. \ref{Fig.4}-b (spatiotemporal and temporal
noise) show these behaviors. Figure \ref{Fig.4}-a shows that there
is a weak dependence of the $SNR_{Global}$ for low values of
$\gamma$. On one hand, in general for the spatiotemporal noise, the
$SNR_{Global}$ is constant and a weak dependence with the $F_{Med}$
is apparent. On the other hand, there exists a dependence of the
$SNR_{Global}$ when temporal noise is applied. Again, we can
distinguish two regimes: a first linear one for low transition rate
($\gamma < 0.5$) and a quasi constant regime for $\gamma>0.5$. The
figure also shows that the SNR response is different when the system
is subject to temporal or spatiotemporal perturbations. It is
important to remark that for large $\gamma$ the temporal SNR is
larger than for spatiotemporal perturbations.

\begin{figure}
\centering
\resizebox{.48\columnwidth}{!}{\includegraphics{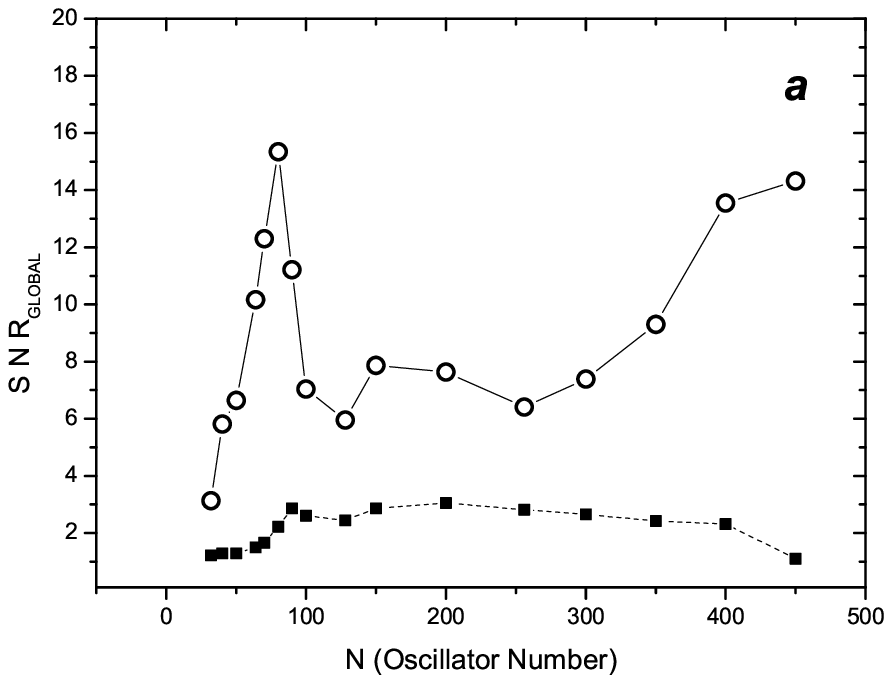}}
\resizebox{.48\columnwidth}{!}{\includegraphics{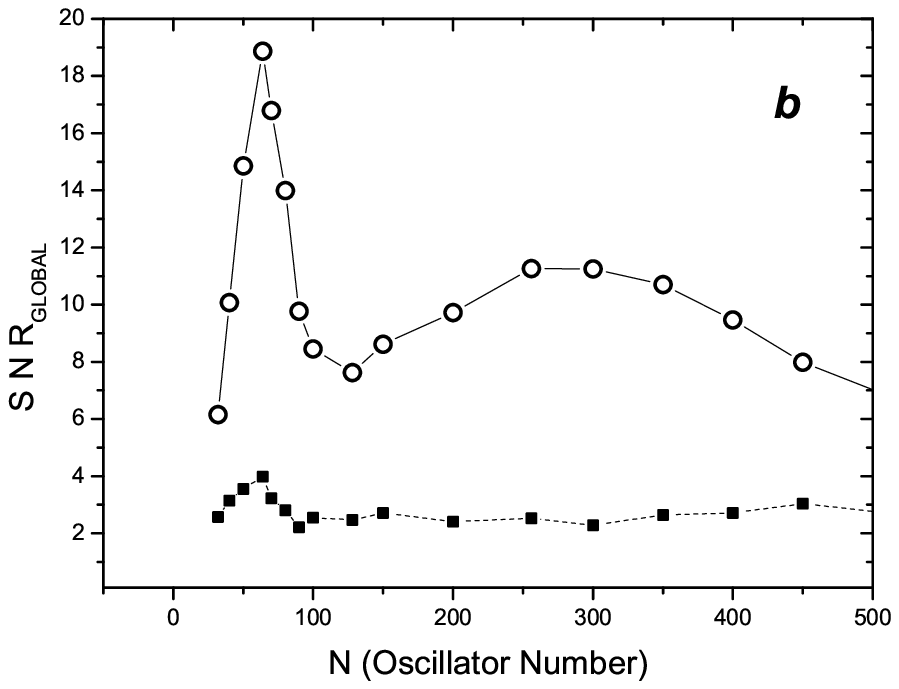}}
\caption{Signal to noise ratio for (a)spatiotemporal evolution . (b)
Temporal evolution. Open circle correspond to $F_{Med}=5$, square to
$F_{Med}=6$. \label{Fig.5}}
\end{figure}

Finally in Figs. \ref{Fig.5}-a and \ref{Fig.5}-b we  show the
results obtained for the behavior of the global SNR as a function of
the system size. The figures show a resonant-like behavior for a
size around $80$ in the low developed chaos case, meanwhile this
phenomenon is weaker when the chaos is more developed. This behavior
is analogous to the so called \textit{system size stochastic
resonance} \cite{sssr}

At this point it is worth to comment on the similarities of the
SR-like phenomena found here and the so called \textit{internal
signal} SR \cite{int}. Previous studies have shown that in some
systems having an internal typical frequency, SR can occur not only
at the frequency of an external driving signal, but also at the
frequency corresponding to the internal periodic behavior
\cite{int}. Regarding the present mechanism of SR, what we can
indeed remark is that the increase in the SNR is related not to a
\textit{reinforcement} of the peak high respect to the noisy
background at a given frequency, but with a \textit{reduction} of
the \textit{pseudo} (or deterministic) noisy background when turning
on the \textit{real} noise. That is, the interplay between ``real"
noise and ``deterministic" noise conforms a kind of
\textit{noise-induced chaos reduction}. Figure 3a shows, for fixed
values of $F$ and $\gamma$, the behavior of $S(\omega)$ in both
cases: with ($\Delta \neq 0$) and without noise ($\Delta = 0$). The
above indicated reduction trend, as the \textit{real} noise is
turned on, is apparent.

The above indicated trend seems to be also responsable of the
behavior observed in Fig. 4a, as $\gamma$ also enters into de
definition of the noise intensity. From Fig. 4b it becomes clear
that the spatial-temporal noise has a stronger influence than the
temporal one on the system response.

\section{Finite perturbations and errors }

The exponential growth of small initial perturbations is the main
effect of chaos and the origin of the lack of prediction in
deterministic dynamical systems. This growth is not homogeneous but
localized in some unstable directions that are fix in low
dimensional cases (usual localization) and moving in the case of
extended chaos (dynamic localization) \cite{PiK}. Moreover, in a
real system perturbations do not grow indefinitely but saturate by
the action of non-linearities. Hence, in real situations we must
deal with finite perturbations \cite{FinPert} that, for small enough
initial perturbations, develop in two regimes, the infinitesimal one
characterized by a exponential growth localized in some directions
and the above mentioned non-linear regime in which saturation by
non-linear effects destroy the exponential growth as well as the
gained localization \cite{Patterns}.

An important problem in predictability analysis is just the
characterization and quantification of both the exponential growth
and the degree of localization. The Lyapunov theory of perturbation
analysis has been a traditional tool to solve particulary the first
part of this problem related with the exponential growth. In the
case of spatial chaos there is a recent theory that accounts for
both parts in a very convenient form. It is based on  mapping
perturbations in rough surfaces trough the application of a
logarithmic transformation (the Hopf-Cole transformation). The use
of this mapping simplifies the analysis of perturbations due to
several reasons; the statistics in the mapped space is Gaussian (or
Poissonian, etc..) instead of being log-normal (log-Poisson, etc..)
\cite{Predict}. Moreover, the growth of rough interfaces is
supported by a well established theory with well defined time and
length scales that are connected by scaling laws \cite{DynScal}, and
finally there are universal types of growth which offers very good
forms of characterization. As we show in the next sections the use
of this mapping provides us with a powerful tool to analyze the
interplay of chaos and noise in a spatial chaotic system.

\subsection{ The Mean-Variance of Logarithms diagram for
perturbations and errors}

{\bf Finite perturbations} from the original Eq.[\ref{Lorenz96_1}]
are obtained by evolving with exactly the same equation (noise
included) a perturbed initial condition $x_{i}'(0)=x_{i}(0)+\delta
x_{i}(0)$. At a longer time finite perturbations are then given by
the difference $\delta x_{i}(t)=x_{i}(t)-x_{i}'(t)$. We refer to it
as a perturbation because only changes in the initial conditions are
considered. In this aspect the noise, which is the same in both
realizations, acts as deterministic and can be considered as a
parametrization of small scale phenomena. On the other hand the
definition of {\bf finite error} is the same
$\epsilon_{i}(t)=x_{i}(t)-x_{i}'(t)$ but now the evolution of both
the control system ($x_{i}(t)$) and the perturbed one ($x_{i}'(t)$)
are obtained with different realizations of noise.

In the infinitesimal regime of finite fluctuations one can write an
equation for perturbations just linearizing around the control
trajectory, that reads
\begin{equation} \label{MVL1}
\frac{d \delta x_{j}}{dt}=x_{j-1}\delta x_{j+1}-x_{j-1}\delta
x_{j-2} + (x_{j+1}-x_{j-2})\delta x_{j-1}-\delta x_{j},
\end{equation}
while for for errors we have
\begin{equation} \label{MVL2}
\frac{d \delta X_{j}}{dt}=X_{j-1}\delta X_{j+1}-X_{j-1}\delta
X_{j-2} + (X_{j+1}-X_{j-2})\delta X_{j-1}-\delta X_{j} + \psi_i (t),
\end{equation}
that is the same equation, but including an additive noise term
$\psi_i (t)= F_{i}(t)-F_{i}'(t)$. Hence, the great difference
between perturbations and errors is that the first have a
multiplicative character while the second include an additive
fluctuation. As we show in the following this is an important fact
in order to reach localization.

Therefore, the multiplicative character of perturbations suggests a
logarithmic transformation $H_{i}(t)=\log(|\delta x_{i}(t)|$ to deal
with more homogeneous relative fluctuations \cite{FinPert}. This can
be achieved by using the Hopf-Cole transformation in Eq.[\ref{MVL1}]
that, considering only the first two terms of the continuous limit
$\frac{\delta x_{j+n}}{\delta x_{j}} \sim 1+n
\partial_x h + \frac{n^2}{2} (\partial_{xx} h + (\partial_{x}
h)^{2}) $ gives
\begin{equation} \label{MVL3}
\partial_{t} H(t)\sim
\xi(x,t)\partial_{x}H(x,t)-\xi(x,t)\frac{1}{2}(\partial_{xx}H
     +(\partial_{x}H)^2)+\eta(x,t).
\end{equation}

We can now interpret the above equation as the growth of a rough
surface $H(x,t)$ with random diffusion and drift
$\xi(j,t)=3x_{j-1}-x_{j+1}+x_{j-2}$, $\eta(j,t)=x_{j+1}-x_{j-2}-1$.
Note that, from this point of view, we are considering the original
field $x(t)$ as an equivalent noise, hence it is the noise generated
by the chaotic system itself, whose stochastic characterization has
been obtained in the previous sections. Hence $\xi$ and $\eta$
become colored noises in space and time. The effect of the external
noise $F(t)$ over the perturbations are indirectly accounted for by
changes in $x(t)$.

Let us now introduce the \textit{Mean-Variance of Logarithms} (MVL)
diagram \cite{MVL} in order to have a graphical representation of
the exponential growth and localization. It is achieved  by
representing the mean value $M(t)=\langle \overline{H(x,t)}\rangle$
and the variance $\langle \overline{h(x,t)^2}\rangle$, ( with
$h(x,t)=H(x,t)-\overline{H(x,t)}$), of $H(x,t)$ (where $< >$ means
average over the ensemble, and $\bar{A}$ corresponds to the space
average), the logarithm of the perturbations. Note that the velocity
of the mapped surface $\lambda=\overline{\dot{H(x,t)}}$ accounts for
the exponential growth since in essence it is the logarithm of the
zero norm of perturbations, namely, the main Lyapunov exponent
\cite{FinPert,Predict}. Hence $M=\lambda t$ is $t $ times the main
Lyapunov exponent. On the other hand we know that the correlation
length of the surface, that evolves as a power law $l_{c}(t)\sim
t^{1/z}$, accounts for the degree of localization, and the variance,
which is the width of the surface, is related with this quantity as
$V(t)\sim l_{c}(t)^{2\alpha}$ \cite{FinPert,Predict}. $z$ and
$\alpha$ are respectively the dynamic and roughness exponents of a
rough interface. They exhibit universal values that in our case (KPZ
universality) are $z=3/2$, $\alpha=1/2$. In summary, depicting
$V(t)$ against $M(t)$ we have a graphical picture of the acquired
localization versus the exponential growth.

\subsection{ Finite perturbations without noise}

\begin{figure}
\centering
\resizebox{.48\columnwidth}{!}{\includegraphics{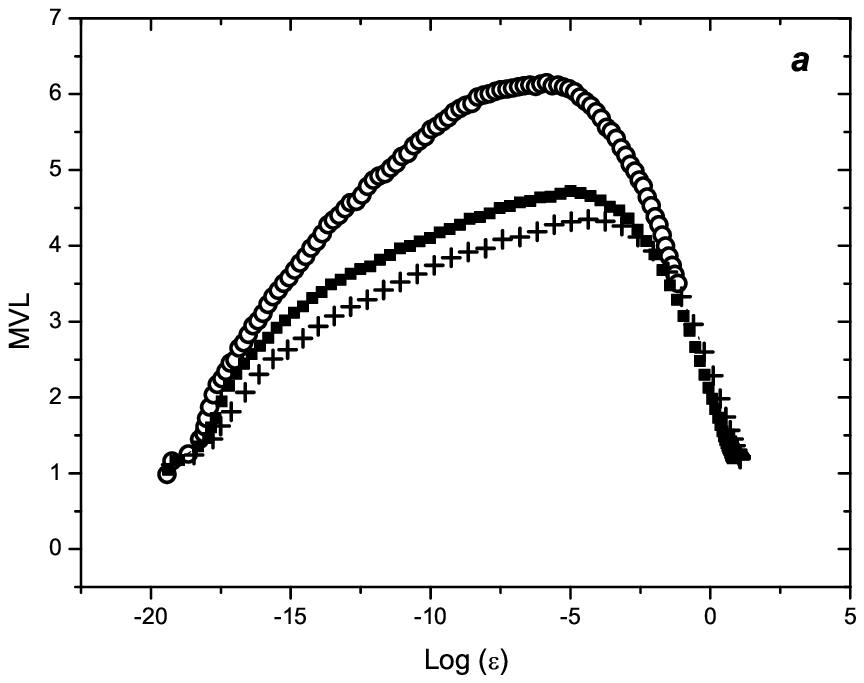}}
\resizebox{.48\columnwidth}{!}{\includegraphics{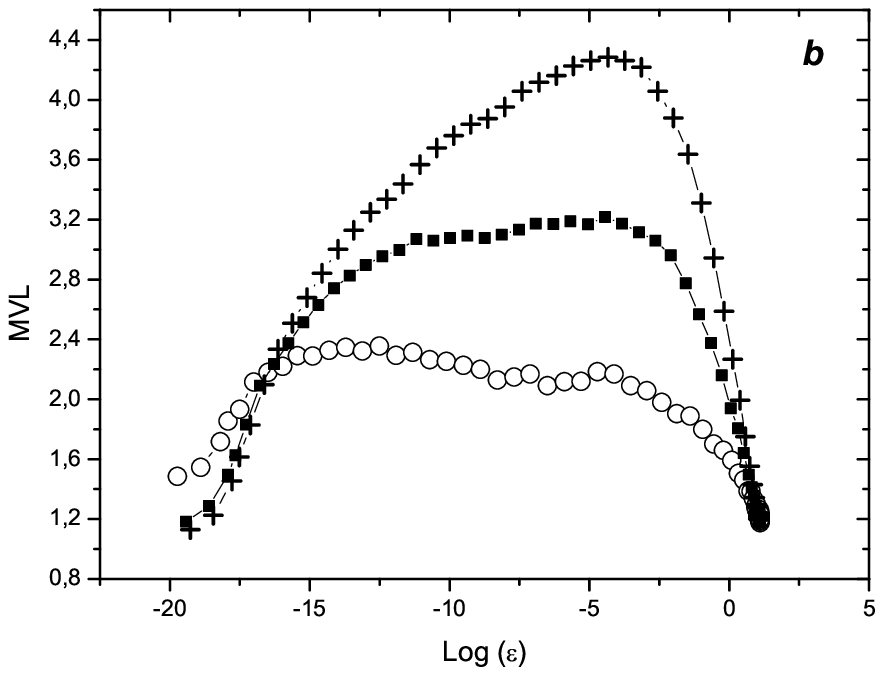}}
\caption{MVL graphics varying (a)$F_{Med}$ parameter, Open circle
$F_{Med}=5$, Square $F_{Med}=6$ and Cross $F_{Med}=8$, for $N=128$ .
(b)System Size. Open circle  $N=32$, Square  $N=64$ and Cross
$N=128$. Common parameters, $\Delta$=0, $Amp=10^{-8}$.
\label{Fig.6}}
\end{figure}

The typical graph of a finite perturbation (see Figs. \ref{Fig.6}-a
and \ref{Fig.6}-b) shows an initial regime corresponding to the
infinitesimal evolution towards the main Lyapunov vector, which
happens increasing spatial correlation and localization, hence we
show an increasing of $\omega^2$ (dispersion), followed by a second
regime where the growth is collapsed by non-linearities and
localization becomes destroyed \cite{Patterns}. We have shown MVL
diagrams in two cases, varying the degree of chaos with the
parameter $F_{Med}$ (Fig. \ref{Fig.6}-a) and varying the system size
$N$.  In the first case we observe that the highest degree of
localization is got for the case of less developed chaos ($\omega^2
\sim 6.5$ with $F_{Med} \sim 5$). Although it is not evident from
intuition it can be expected since in this case the intensity of the
deterministic noise, given by the area of the spectrum of $\xi$, is
greater than in the case of more developed chaos. It is worth
observing the high level of localization obtained, $\omega^2 \sim
6$, in this case. Despite of being a case of low developed chaos the
effect on spatial propagation of perturbations is very strong. Note
that in all cases in this figure infinitesimal fluctuations saturate
by non-linearities since we are dealing with an enough large system,
$N=128$. By contrast, in the Fig. \ref{Fig.6}-b we show how with
small systems saturation due to finite size takes place. With very
small size, $N=32$, saturation by finite size only allows a small
localization ($\omega^2 \sim 2$), that grows for larger systems,
$\omega^2 \sim
 3$ for $N=64$ and $\omega^2\sim 4$ for $N=128$. This fact can also
be expected from the mapping to rough interfaces. The width of a
rough interface in the KPZ universality scales with the system size
as $\omega^2 \sim N$. This scaling …

\subsection{ Localization of perturbations and errors}

\begin{figure}
\centering
\resizebox{.48\columnwidth}{!}{\includegraphics{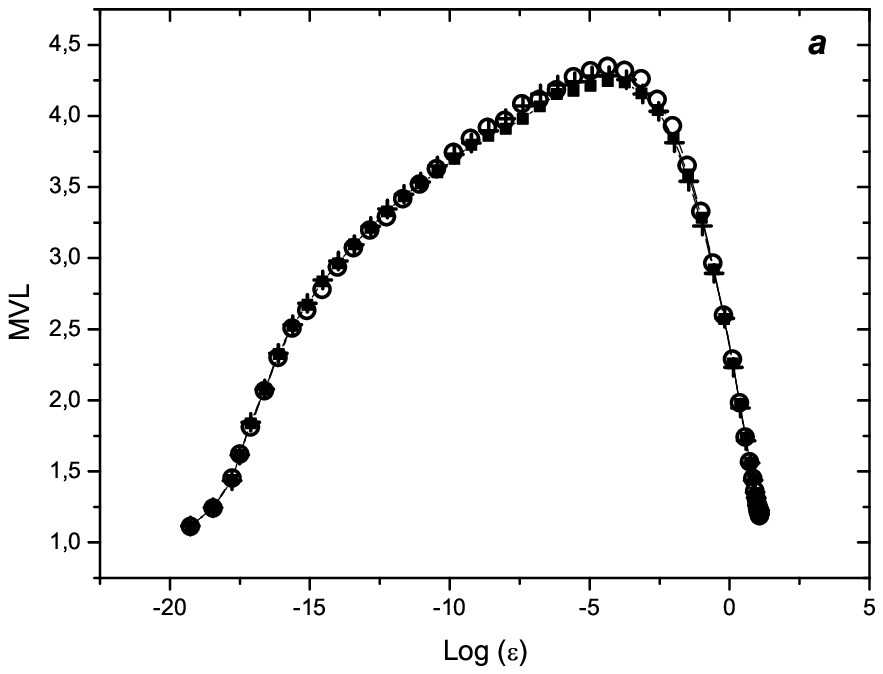}}
\resizebox{.48\columnwidth}{!}{\includegraphics{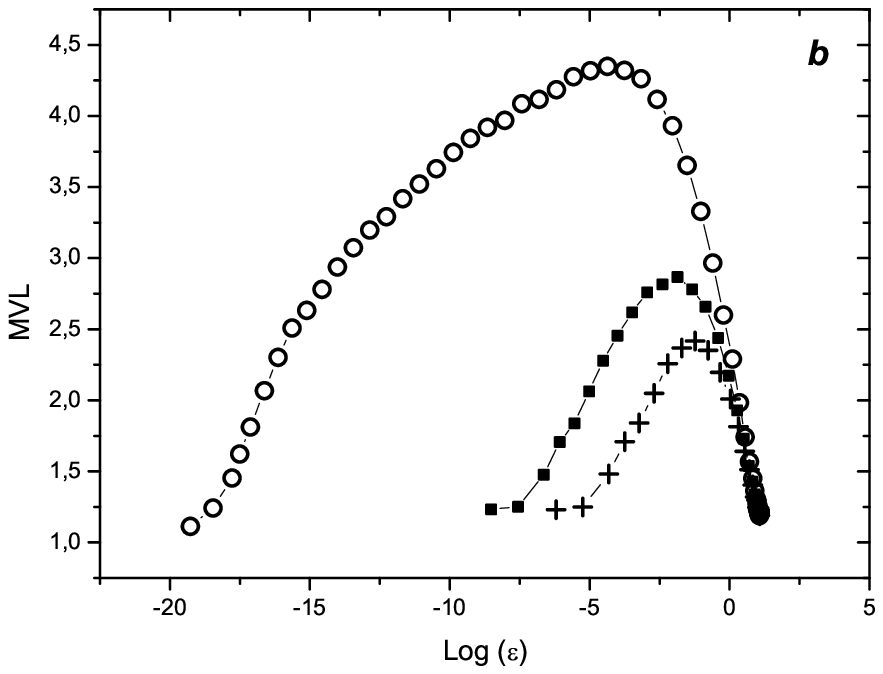}}
\caption{MVL graphics for (a) perturbations evolution. (b) errors
evolution. Open circle corresponds to non-noise evolution
$\Delta=0$, square to $\Delta=0.001$ and cross to $\Delta=0.01$.
Common parameters, $N=128$, $F_{Med}=8.0$, $Amp=10^{-8}$.
\label{Fig.7}}
\end{figure}

In the   figure 7 (see Figs. \ref{Fig.7}-a and \ref{Fig.7}-b)we show
the MVL diagram of (a) perturbations and (b) errors for distinct
values of the noise amplitude. In a first look we observe that the
effect of noise seems to be irrelevant in the case of perturbations
(see Fig. \ref{Fig.7}-a) but it is very important in the case of
errors (see Fig. \ref{Fig.7}-b). This is an interesting result that
shows the differences between multiplicative and additive
fluctuations. In the case of perturbations the external noise keeps
the multiplicative character of the evolution of perturbations
(Eq.[\ref{MVL1}) and it only acts changing the equivalent
deterministic noise trough $\xi(t)$. As a consequence the evolution
of perturbations results slight affected by noise. On one hand in
Fig. \ref{Fig.7}-a we can see a very high level of localization
$\omega^2 > 4$ in all cases. On the other hand, errors evolve with
the external noise as an additive fluctuation. Then, a competition
between the deterministic multiplicative noise, that tries to
localize the error (increasing of $\omega^2$), against the additive
external fluctuation that does not produce localization, occurs as
observed in the figure. Important changes in localization,
$\omega^2=4,3,2.5$ for distinct noise amplitudes
($\Delta=0,0.001,0.01$) are shown in this figure.

\section{\label{conc}Conclusions}

We have investigated the effect of a time-correlated noise on an
extended chaotic system, analyzing the competence between the
indicated \textit{deterministic} or \textit{pseudo-noise} and the
real random process. For our study we have chosen the Lorenz'96
model \cite{Lor96-1} that, in spite of the fact that it is a kind of
\textit{toy} model, is of interest for the analysis of climate
behavior \cite{Lor96-2,Lor96-3}. It is worth remarking that it
accounts in a simple way for the spatial structure of geostrophic
waves and the dynamics of tropical winds. The time series obtained
at a generic site $x_i(t)$ mimics the passing of such waves, which
is in fact a typical forecast event. We have assumed that the unique
model parameter $F$ is time dependent and composed of two parts, a
constant deterministic, and a stochastic contribution in both
temporal and spatiotemporal forms.

We have done a thorough analysis of the system's temporal evolution
and its time correlations. The action of a stochastic noise on the
Lorenz'96 system produces a dissipation on the time evolution. This
dissipation essentially depends on the $F_{Med}$ parameter.
Furthermore our results show numerical evidence for two SR-like
behaviors. In one hand a ``normal" SR phenomenon that occurs at
frequencies that seems to correspond to a system's quasi-periodic
behavior. On the other hand, we have found a SSSR-like behavior,
indicating that there is an optimal system size for the analysis of
the spatial system's response. As indicated before, the effect of
noise is stronger when the chaos is underdeveloped.

We argue that these findings are of interest for an \textit{optimal}
climate prediction. It is clear that the inclusion of the effect of
an external noise, that is a stochastic parametrization of unknown
external influences, could strongly affect the deterministic system
response, particularly through the possibility of an enhanced
system's response in the form of resonant-like behavior. It is worth
here remarking the excellent agreement between the resonant
frequencies and wave length found here, and the estimates of Lorenz
\cite{Lor96-1p}.

The effect of noise is weak respect to changes in the spatial
structure, with the main frequencies remaining unaltered, but it is
strong concerning the strength of the ``self-generated"
deterministic noise. We expect that in such a system and at the
resonant frequencies, forecasting would be improved by the external
noise due to the effect of suppression of the self-generated chaotic
noise. Such an improvement will become apparent through the analysis
of the localization phenomenon in the MVL diagrams. The detailed
analysis of such an aspect will be the subject of a forthcoming
study \cite{nos1}.\\

\textbf{Acknowledgments:} We acknowledge financial support from MEC,
Spain, through Grant No. CGL2004-02652/CLI. JAR thanks the MEC,
Spain, for the award of a \textit{Juan de la Cierva} fellowship. HSW
thanks to the European Commission for the award of a \textit{Marie
Curie Chair} during part of the development of this work.

\end{document}